\documentstyle[12pt]{article}

\catcode `\@=11 \@addtoreset{equation}{section}

\catcode `\@=12

\newcommand{\be}{\begin{equation}}
\newcommand{\en}{\end{equation}}
\newcommand{\bea}{\begin{eqnarray}}
\newcommand{\ena}{\end{eqnarray}}
\newcommand{\beano}{\begin{eqnarray*}}
\newcommand{\enano}{\end{eqnarray*}}
\newcommand{\bee}{\begin{enumerate}}
\newcommand{\ene}{\end{enumerate}}

\newcommand{\R}{R \!\!\!\! R}
\newcommand{\N}{N \!\!\!\!\! N}
\newcommand{\Z}{Z \!\!\!\!\! Z}

\newcommand{\Lc}{{\cal L}}
\newcommand{\D}{{\cal D}}
\newcommand{\C}{{\cal C}}

\begin{document}

\thispagestyle{empty}

\vspace*{1cm}

\begin{center}
{\Large \bf Multi-Resolution Analysis and Fractional Quantum Hall Effect: More Results}   \vspace{2cm}\\

{\large F. Bagarello}
\vspace{3mm}\\
  Dipartimento di Matematica ed Applicazioni,
Fac. Ingegneria, Universit\`a di Palermo, Viale delle Scienze, I-90128  Palermo, Italy\\
e-mail: bagarell@unipa.it
\vspace{2mm}\\
\end{center}

\vspace*{2cm}

\begin{abstract}
\noindent In a previous paper we have proven that any
multi-resolution analysis of $\Lc^2(\R)$  produces, for even
values of the inverse filling factor and for a square lattice, a
single-electron wave function of the lowest Landau level (LLL)
which, together with its (magnetic) translated, gives rise to an
orthonormal set in the LLL. We have also discussed the inverse
construction.

In this paper we simplify the procedure, clarifying the role of
the kq-representation. Moreover, we  extend our previous results
to the more physically relevant case of a triangular lattice and to
odd values of the inverse filling factor.  We also comment on
other possible shapes of the lattice as well as on the extension
to other Landau levels.

Finally, just as a first  application of our technique, we
compute (an approximation of) the Coulomb energy for the Haar wavefunction, for a
filling $\nu=\frac{1}{3}$.
\end{abstract}

\vspace{1.7cm}


{\bf Running title}:  Multi-Resolution Analysis and Fractional Quantum Hall Effect

\vfill

\newpage

\section{Introduction}

Since its discovery, the fractional quantum Hall effect (FQHE) has been considered as a very exciting problem to be understood. For this reason, it is not surprising that the presence of the plateaux in the Hall resistivity has been explained in many different ways by many different people. Most of these proposals are analyzed in monographies like \cite{gir2,dassarma,gir}. The current belief is that the Laughlin wave function $\Psi_L$ can explain the existence of these plateaux. As it is known, $\Psi_L$
corresponds to an incompressible fluid and it describes electrons which are strongly correlated. It is also well know, however, that for certain values of the filling factor $\nu$, the electrons behave like a two-dimensional crystal, so that they must be described by a sustantially different wave function, $\Psi^{(N)}$, which is usually taken as a normalized Slater determinant constructed starting with single electron normalized wave functions which should be mutually orthogonal in order $\Psi^{(N)}$ to be normalized in the thermodynamical limit, \cite{bms}. Many proposals in this direction has been made during these years, \cite{yos,lam,maki,yos2,hald,ferr} among the others, most of which are unable to reproduce the critical value of the filling factor for which a transition from liquid to crystal has been observed, \cite{lev,gold,zhu}.

In this paper we are mainly interested to those values of $\nu$ for which the electrons behave as a 2-dimensional crystal. This aspect of the FQHE has already been considered in many papers, \cite{yos,maki,hald} etc.. What is new here is the use of a technique coming from wavelet theory, the multi-resolution analysis (MRA) of $\Lc^2(\R)$, which appears to be rather promising, as we will show in this work.

In a previous paper, \cite{mra1}, we have  proven that for a
two-dimensional electron gas (2DEG) in a strong orthogonal
magnetic field there exists a deep relation between any
MRA of $\Lc^2(\R)$  and a normalized
wave function for the $N$ electrons system in the lowest Landau level (LLL).
In particular we have seen that any MRA produces a single electron
wave function which, together with its (magnetic) translated,
produces an orthonormal set in the LLL. We have also proved that
the procedure can be inverted. This surprising result, however,
was obtained assuming that the electrons live into a square lattice and, more important,
was shown to hold only for a filling factor $\nu=\frac{1}{2L}$,
$L\in\N$. These mathematical constraints are physically rather
unpleasant. In fact, classical and quantum arguments suggest that
for low density the electrons should arrange themselves into a
triangular lattice, \cite{bon,bms}, rather than into a square one.
This is due to the fact that the triangular lattice minimizes, for
a fixed electron density, the Coulomb energy. Therefore, in order
to have a deeper understanding about the concrete relevance of
wavelets in FQHE, it is absolutely necessary to extend the results
 in \cite{mra1} to the triangular lattice. Moreover, the
analysis of this new geometry, gives a new insight about the
procedure and, in particular, clarifies the role of the
$kq$-representation, which was an essential mathematical tool in our original approach.

This is not the end of the story. It is well known that the main
plateaux in the experimental measures of the resistivity tensor
appear for filling factors like
$\nu=\frac{1}{3},\frac{1}{5},\frac{1}{7},..$. This fact seems to suggest that our technique, \cite{mra1}, is
useless in most physically relevant situations (even if fillings like $1/2n$ have a physical interest by their own right!).

In this paper we overcome both these limitations: we show first
how to obtain an orthonormality condition for the single electron
wave functions for the triangular lattice. We also show that the
same procedure can be adapted to lattices of arbitrary shapes,
provided that the so-called rationality condition is satisfied.
Incidentally, this extension reduces the relevance of the
$kq$-representation with respect to \cite{mra1}.

We also prove another  result which looks quite fascinating to us:
the odd values of $\nu^{-1}$ are recovered by MRA of a
different kind, that is MRA with dilation parameter $d$ different
from 2. In particular we will prove that a $d$-MRA, $d$ natural
and bigger than 2, produces an orthonormal set of single electron wave functions in the LLL related
to a filling $\nu=\frac{1}{d}$. The extension to other Landau
levels is also discussed in some details, as well as the
possibility of reversing the construction.

The paper ends with an example of a 3-MRA which extends the usual
Haar construction, \cite{dau}, which is used to produce an
$N$-electrons Slater determinant made up with orthonormal single
electron wave functions, for $\nu=\frac{1}{3}$. Finally, we
compute the quantum Coulomb energy associated to this state under
some simplifying approximations.

\section{Mathematical tools}

In order to keep the paper self-contained we now quickly review,
for reader's convenience, the main definitions of the mathematical
tools we will use in the rest of the paper.

\subsection{$d$-Multi-Resolution Analysis}

The main result in the theory of $d$-MRA is the recipe which
allows to construct an orthonormal basis in $\Lc^2(\R)$ starting
from a set of functions $\psi^{(i)}$, $i=1,2,...,d-1$, and acting
on these with dilation and translation operators.

The full story for $d=2$ may be found, for instance, in
\cite{dau}, which also contains some  remarks concerning $d=3$.
For a more complete reading on this subject we suggest references
\cite{stri,gro}, where the definition of MRA is extended to
$\Lc^2(\R^n)$ and the dilations and translations are not
necessarily related to the integer numbers. Here, however, we give
the definition of a $d$-MRA for $\Lc^2(\R)$ and $d$ integer (and
$d\geq 2$), and then we focus only on those aspects which will be
useful in the following.

A $d$-{\em multi-resolution analysis} of $L^2(\R)$ is an
increasing sequence of closed subspaces \be \ldots \subset V_{-2}
\subset V_{-1} \subset V_0 \subset V_1 \subset V_2
  \subset \ldots,
\en with $ \bigcup_{j \in \Z} V_j$ dense in $\Lc^2(\R)$ and $
\bigcap_{j \in \Z} V_j = \{0\}$, and such that
\begin{itemize}
\item[(1)]
$f(x) \in  V_j \Leftrightarrow f(dx) \in  V_{j+1}$
\item[(2)]
There exists a function $\phi \in V_0$, called a  {\em scaling}
function, such that  $\{\phi(x-k), k \in {\Z}\}$ is an o.n. basis
of $V_0$.
\end{itemize}
Combining (1) and (2), one gets an  o.n. basis of $V_j$, namely $
\{\phi_{j,k}(x) \equiv d^{j/2} \phi(d^jx-k),$ $ k \in \Z \}.$

For an integer $d\geq 2$  the existence of a set of functions
$\psi^{(i)}$, $i=1,2,...,d-1$, such that $\{\psi_{j,k}^{(i)}(x)
\equiv d^{j/2} \psi^{(i)}(d^jx-k), j,k \in \Z\,\: i=1,2,...,d-1\}$
constitutes an o.n. basis of $\Lc^2(\R)$ can be proved. We are not
interested in all the details of this construction here, see references \cite{stri,gro}, but only to some easy
consequences of this definition.

Notice that the inclusion $V_0 \subset V_1$ yields the relation
\be   \label{phi} \phi(x) = \sqrt{d} \,\sum_{n=-\infty}^\infty \;
h_n \phi(dx - n), \quad h_n = \langle \phi_{1,n} | \phi
\rangle=\sqrt{d}\int_{\R}\overline{\phi(dx-n)}\phi(x)dx. \en The
orthonormality requirement of the functions $\phi$,
$\int_{\R}\overline{\phi(x-n)}\phi(x)dx=\delta_{n,0}$, together
with the equation (\ref{phi}), produces the key equation for the
coefficients ${h_n}$: \be
\sum_{n\in\Z}h_n\overline{h_{n+dl}}=\delta_{l,0}, \label{21} \en
which generalizes equation (2.12) of \cite{mra1} to a dilation
parameter $d\geq 2$.

It is also easily checked that, if $\int_{\R}\phi(x)dx\neq0$, then
the coefficients must also satisfy the sum rule \be
\sum_{n\in\Z}h_n=\sqrt{d}. \label{22} \en

Equation (\ref{21}) is all what we will need in the following. It
is clear from the above construction that any $d$-MRA produces a
set of complex quantities $h_n$ which satisfies this condition.
The converse procedure, which is not so interesting for our aims,
is discussed in \cite{stri} and requires some care: in particular,
it is easy to produce examples, for $d\geq 3$, of $\{h_n\}$
satisfying condition (\ref{21}) and a {\em scaling function}
$\phi$ satisfying the $d$-scale relation (\ref{phi}), which do not
produce a $d$-MRA.

\subsection{kq-representation}

We give here only few definitions, namely those relevant for this
paper, and refer to \cite{zak,jan,zak2}  for further reading and
for applications.

The genesis of the kq-representation lays in the well known
possibility of a simultaneous diagonalization of any two commuting
operators. In particular, the following distributions \be
\psi_{kq}(x)=\sqrt{\frac{2\pi}{a}}
\sum_{n\in\Z}e^{ikna}\delta(x-q-na), \quad\quad k\in[0,a[, \quad
q\in[0,\frac{2\pi}{a}[ \label{26} \en are (generalized)
eigenstates of both $T(a)=e^{ipa}$ and
$\tau(\frac{2\pi}{a})=e^{ix2\pi/a}$. Here $a$ is a given positive
real number.

These $\psi_{kq}(x)$ are Bloch-like functions corresponding to
infinitely localized Wannier functions, \cite{zak2}. They also
satisfy orthogonality and closure properties. This implies that,
roughly speaking, they can be used to define a new representation
of the wave functions by means of the integral transform $Z:
\Lc^2(\R)\rightarrow \Lc^2(\D)$,
$\D=[0,a[\times[0,\frac{2\pi}{a}[$, defined as follows: \be
h(k,q):=(ZH)(k,q):=\int_{\R}d\omega
\overline{\psi_{kq}(\omega)}H(\omega), \label{27} \en for any
given function $H(\omega)\in \Lc^2(\R)$. The result is a function
$h(k,q)\in \Lc^2(\D)$.

To be more rigorous, $Z$ should be defined first on the functions
of ${\cal C}_o^\infty(\R)$ and then extended to $\Lc^2(\R)$ using
the continuity of the map, \cite{dau2}.

Replacing  $\psi_{kq}(x)$ with its explicit expression, formula
(\ref{27}) produces \be
h(k,q)=(ZH)(k,q)=\sqrt{\frac{2\pi}{a}}\sum_{n\in
\Z}e^{-ikna}H(q+na), \label{29} \en which can be inverted and
gives the $x$-representation $H(\omega)\in \Lc^2(\R)$ of a
function $h(k,q)\in \Lc^2(\D)$ as follows: \be
H(\omega)=(Z^{-1}h)(\omega)=\int_{\D}dk \, dq
\psi_{kq}(\omega)h(k,q). \label{210} \en Due to  (\ref{26}), this
equation can be written as \be
H(\omega+na)=\sqrt{\frac{2\pi}{a}}\int_0^b dk e^{ikna}
h(k,\omega), \quad \forall \omega\in [0,a[, \, \forall n\in \Z.
\label{212} \en

\section{Stating the problem}

The physical system we are considering is a  2DEG, that is a gas
of electrons constrained in a two-dimensional layer belonging to
the $(O;x,y)$ plane, in a neutralizing positive uniform background
and subjected to an uniform magnetic field along $z$.

The hamiltonian of the $N$-electrons system can be written as
\begin{equation}
H^{(N)}=H^{(N)}_0+\lambda(H^{(N)}_c+H^{(N)}_B)\label{31}
\end{equation}
where $H^{(N)}_0$ is the sum of $N$ contributions:
\begin{equation}
H^{(N)}_0=\sum^N_{i=1}H_0(i).\label{32}
\end{equation}
Here $H_0(i)$ describes the minimal coupling of the electrons with
the magnetic field:
\begin{equation}
H_0={1\over 2}\,\left(\underline p+\underline
A(r)\right)^2={1\over 2}\,\left(p_x-{y\over 2}\right)^2+{1\over
2}\,\left(p_y+{x\over 2}\right)^2. \label{33}
\end{equation}
Notice that we are adopting here the symmetric gauge $\underline
A=\frac{1}{2}(-y,x,0)$ and the same units as in \cite{bms}.
$H^{(N)}_c$ is the canonical Coulomb interaction between charged
particles:
\begin{equation}
H^{(N)}_c={1\over2}\,\sum^N_{i\not=j}{1 \over|\underline r_i-
\underline r_j|}\label{34}
\end{equation}
and $H^{(N)}_B$ is the interaction of the charges with the
background, whose explicit form is irrelevant here and can be found in \cite{bms}.

In the following we will consider, as it is usually done in the
literature, $\lambda(H^{(N)}_c+H^{(N)}_B)$ as a perturbation of
the free hamiltonian $H^{(N)}_0$, and we will look for eigenstates
of $H^{(N)}_0$ in the form of Slater determinants built up with
single electron wave functions. This approach is known to give
good results for low electron (or hole) densities, \cite{maki,bms}.
However, comparison with the experiments shows also that the
wave-functions proposed in \cite{bms,maki,yos,ferr}  does not reproduce completely
the  experimental data, as far as the transition from the Wigner
crystal to the Laughlin liquid is concerned. Other wave-functions
proposed in the literature share with this the same problem, so
that the problem of finding the correct single-electron wave
function describing the system for low electron density is still
open. In this paper we will show how an o.n. set of wave functions
in the LLL (as well as in any other Landau level) can be
constructed easily starting from a $d$-MRA.

The easiest way to attach this problem consists in introducing the
new variables
  \be
\label{35}
  P'= p_x-y/2, \hspace{5mm}     Q'= p_y+x/2.
  \en
In terms of $P'$ and $Q'$ the single electron hamiltonian, $H_0$,
can be written as
 \be
\label{36}
  H_{0}=\frac{1}{2}(Q'^2 + P'^2).
  \en
The transformation (\ref{35}) can be seen as a part of a canonical
map $U$ from $(x,y,p_x,p_y)$ into $(Q,P,Q',P')$ where

   \be
\label{37}
   P= p_y-x/2+\frac{1}{\sqrt{3}}(p_x+y/2), \hspace{5mm}
  Q= p_x+y/2.
   \en
  These operators  satisfy the following commutation relations:
  \be
\label{38}
 [Q,P] = [Q',P']=i, \quad  [Q,P']=[Q',P]=[Q,Q']=[P,P']=0.
  \en
   Using \cite{mo}, we deduced in \cite{bagphd}  that, as a consequence of
   the above canonical transformation, a wave function in the $(x,y)$-space
is
   related to its  expression in terms of the new variables $(P,P')$  by the formula
  \be
\label{39}
  \psi(x,y)=\frac{e^{\frac{iy}{2}(x-\frac{y}{\sqrt{3}})}}{2\pi}\int_{-\infty}^{\infty}\,
  \int_{-\infty}^{\infty}e^{i\{P'(x-\frac{y}{\sqrt{3}})+yP+PP'-\frac{P'^2}{2\sqrt{3}}\}}\psi(P,P')\,dP \,dP'.
  \en
We want to stress that this formula differs from the analogous one
in \cite{mra1} since the map $U:(x,y,p_x,p_y)\rightarrow
(Q,P,Q',P')$ differs from the one used in \cite{mra1}. The reason
for this difference is in the shape of the lattice, which produces
a different canonical transformation. This will appear  clearly in
the following, when we will discuss how the lattice should be
constructed explicitly and the role of the magnetic translations
in this construction.

  The usefulness of the  new variables stems from the expression
(\ref{36})
  of $H_0$. Indeed, in this representation, the single electron Schr\"{o}dinger equation
admits
  eigenvectors $\psi(P,P')$ of $H_0$ of the form $\psi(P,P')=f(P')h(P)$.
Thus
   the ground state of (\ref{36}) must have the form
 $f_0(P')h(P)$,  where
   \be     \label{310}
  f_0(P')= \pi^{-1/4} e^{-P'^2/2},
  \en
  and the function $h(P)$ is arbitrary in $\Lc^2(\R)$, which manifests the degeneracy of
the LLL. With $f_0$ as above formula (\ref{39}) becomes \be
 \label{311}
  \psi(x,y) =
\int_{-\infty}^{\infty}\,K_0^{(t)}(\underline r, s)h(s)\,ds,
 \en
where $K_0^{(t)}$ is what we call the kernel of the
transformation, \be K_0^{(t)}(\underline r, s)=
\frac{e^{\frac{iy}{2}(x-\frac{y}{\sqrt{3}})}}{\pi^{3/4}\sqrt{2(1+\frac{i}{\sqrt{3}})}}
  e^{iys-\beta(x-\frac{y}{\sqrt{3}}+s)^2},
\label{311bis} \en where
$\beta=\frac{3}{8}(1-\frac{i}{\sqrt{3}})$. Here we use the suffix
{\em t} to emphasize the shape of the lattice, triangular in
this paper, while the index 0 means that we are working in the
LLL. Just to compare this result with the one obtained for the
square lattice, \cite{mra1}, we recall that $K_0^{(s)}(\underline r,
s)=\frac{e^{ixy/2}}{\sqrt{2}\pi^{3/4}} e^{iys}e^{-(x+s)^2/2}$. The
extensions to higher Landau levels (see again \cite{mra1} for
$K_1^{(s)}$), can be found simply replacing $f_0$ in (\ref{310})
with the excited harmonic oscillator eigenstates: using $f_l$
will produce, clearly, a wave function in the l-th Landau level.
However, for large magnetic fields, it is well known that only the
very first levels are relevant in the analysis of the FQHE, and
this is the main reason why, quite often, only the LLL is
considered. This is also our choice here, and for this reason, along this paper, we
will  consider only $K_0^{(t)}$,  but for few
comments.

Once the generic form of any function in the LLL has been given,
 formula (\ref{311}), it remains to choose properly one of
them, that is,  to fix $h(s)$ by requiring that it minimizes the
Coulomb interaction. To achieve this aim we will first construct a particular ground state of the (infinitely degenerate) free $N$-electrons hamiltonian
$H^{(N)}_0$. We use a suggestion coming from the classical
counterpart of this quantum problem. It is very well known that
the ground state for  a classical 2DEG is  a (triangular) Wigner
crystal: the classical electrons are sharply localized on the
sites of a  lattice whose lattice spacing is fixed by the electron
density, \cite{bon}. What we expect, and what was proven in \cite{bms}, is
that, at least for certain values of the filling factor, the
quantum ground state should not be very different from this
classical picture.

Let us introduce the so-called magnetic translation operators
$T(\vec{a_i})$ defined by
   \be
  T(\vec{a_i})\equiv\exp{(i\vec{\Pi}_c\cdot\vec{a_i})},
  \;\;i=1,2,
\label{312}
  \en
  where $\vec{\Pi}_c\equiv(Q,P)$ and $\vec{a_i}$ are the lattice
   basis vectors:
\be \vec{a_1}=a(1,0),\quad \vec{a_2}=\frac{a}{2}(1,\sqrt{3}).
\label{312bis} \en

We assume the following (minimal) {\em rationality condition} on
the area of the fundamental cell of the lattice: \be
a_{1_x}a_{2_y}-a_{1_y}a_{2_x}=2\pi, \label{313} \en which for the
triangular lattice reads \be a^2=\frac{4\pi}{\sqrt{3}}.
\label{315bis} \en

Notice that this is different from the analogous condition for a
square lattice, $a^2=2\pi$, \cite{mra1}. Because of this condition
we have \be [T(\vec{a_1}),T(\vec{a_2})]=0, \label{314} \en which,
together with \be [T(\vec{a_1}), H_0]=[T(\vec{a_2}),H_0]=0,
\label{314bis} \en which easily follow from (\ref{38}), shows that
$T(\vec{a_1}), T(\vec{a_2})$ and $H_0$ are a set of commuting
operators.

This implies that if $\psi$ is an eigenstate of $H_0$ with
eigenvalue $\epsilon$, then $T(\vec{a_1})^nT(\vec{a_2})^m\psi$ is
still an eigenstate of $H_0$ corresponding to the same eigenvalue,
for all $n,m\in \Z$. This can be seen as another evidence of the
infinite degeneracy of the different Landau levels.

Because of (\ref{312bis}),
the magnetic translations take the form \be
T_1:=T(\vec{a_1})=e^{ia(p_x+\frac{y}{2})}=e^{iaQ}, \quad
T_2:=T(\vec{a_2})=e^{i\frac{a}{2}((p_x+y/2)+\sqrt{3}(p_y-x/2))}=e^{i\frac{a\sqrt{3}}{2}P}=e^{i
\frac{2\pi}{a}P}, \label{315} \en which show the relevance of the
definitions in (\ref{37}). Notice that the last equality for $T_2$ above is a simple
consequence of the rationality condition (\ref{315bis}).

Using this equality we can check that the action of $T_1$ and
$T_2$ on a generic function $f(x,y)\in \Lc^2(\R^2)$ is given by
\be f_{m,n}(x,y):=T_1^mT_2^n
f(x,y)=(-1)^{mn}e^{\frac{i}{2}(Y_{nm}x-X_{nm}y)}f(x-X_{nm},y-Y_{nm}).
\label{316} \en

We see from this formula that if, for instance, $f(x,y)$ is
localized around the origin, then $f_{m,n}(x,y)$ is localized
around the lattice site $(X_{nm},Y_{nm})$. Here
$X_{nm}=-a(n+\frac{m}{2})$ and
$Y_{nm}=-am\frac{\sqrt{3}}{2}=-m\frac{2\pi}{a}$. This result
extends, as  expected, the one for the square lattice,
\cite{mra1}, and explains why we call $T_j$ {\em translation operators}.

Moreover, equation (\ref{316}) is the key relation by means the quantum triangular lattice can be constructed: we simply start
from the single electron ground state of $H_0$ given in
(\ref{311}), $\psi(x,y)$. Then we construct a set of copies
$\psi_{m,n}(x,y)$ of $\psi$ as in (\ref{316}), with $m,n\in\Z$.
All these functions still belong to the lowest Landau level for
any choice of the function $h(P)$, due to (\ref{314bis}), and are localized around the sites of our triangular lattice. $N$ of
these wave functions $\psi_{m,n}(x,y)$ are finally used to
construct a Slater determinant for the finite system: \be
\psi^{(N)}(\underline r_1,\underline r_2,...,\underline r_N)= \!\!
\frac{1}{\sqrt{N!}}\left|
\begin{array}{ccccccc}
\psi_{m_1,n_1}(\underline r_1) & \psi_{m_1,n_1}(\underline r_2) & . & . & . & . & \psi_{m_1,n_1}(\underline r_N)  \\
\psi_{m_2,n_2}(\underline r_1) & \psi_{m_2,n_2}(\underline r_2) & . & . & . & . & \psi_{m_2,n_2}(\underline r_N)  \\
. & . & . & . & . & . & .   \\
. & . & . & . & . & . & .   \\
. & . & . & . & . & . & .   \\
\psi_{m_N,n_N}(\underline r_1) & \psi_{m_N,n_N}(\underline r_2) & . & . & . & . & \psi_{m_N,n_N}(\underline r_N)  \\
\end{array}
\right| \label{317} \en We have already discussed in \cite{mra1}
the relevance of the ortonormality of differently localized single
electron wave functions when the thermodynamical limit of the
model is considered. For this reason we look for o.n. single
electron wave-functions. Therefore, let $\psi(x,y)$ be as in
(\ref{311}) and $\psi_{n,m}(x,y)=T_1^nT_2^m\psi(x,y)$. We are
interested  in finding  conditions on $h(P)$ such that the
orthonormality condition (ONC) \be
S_{n,m}:=\int_{\R^2}\overline{\psi_{0,0}(x,y)}\psi_{n,m}(x,y)dx
dy=\delta_{m,0}\delta_{n,0}, \label{319} \en is satisfied. Any
function $\psi(x,y)$ which satisfies this condition and which belongs to the LLL can be used to
generate the function $\psi^{(N)}(\underline r_1,\underline
r_2,...,\underline r_N)$ as in (\ref{317}). Moreover, it is not
hard to prove that the ONC can be rewritten as \be
S_{n,m}:=\int_{\R} ds e^{ism\frac{2\pi}{a}}\overline{h(s)}
h(s-na)=\delta_{m,0}\delta_{n,0}. \label{320} \en This is a
consequence of eqs. (\ref{311}), (\ref{311bis}), (\ref{315bis})
and (\ref{316}).

It is interesting to remark that this equation implies that
$\psi_{0,0}$ is normalized in $\Lc^2(\R^2)$ if and only if $h(s)$
is normalized in  $\Lc^2(\R)$. This is clearly a consequence of
the canonicity of the map $U$.

Moreover, equation (\ref{320}) coincides exactly with the
analogous one we have obtained for the square lattice, under the
condition $a^2=2\pi$, \cite{mra1}. Incidentally, this fact
strongly suggests that the same condition will appear
independently of the shape of the lattice. Finally, we observe
that the ONC (\ref{319}) produces the same equation (\ref{320})
independently of the particular Landau level we are considering.
In other words, it is not hard to check that we obtain  equation (\ref{320}) even if the kernel
$K_0^{(t)}$ is replaced by $K_l^{(t)}$, for any given $l\in \N$.
This proves, therefore, that the problem of the orthonormality of the single electron wave functions is
{\em Landau-level independent}. Of course, this does not means
that the o.n. wavefunctions look the same in any Landau level:
this is true only at the level of the function $h(s)$. On the
contrary, the wavefunctions in the coordinate representation will be
different because the kernels of the transformations,
$K_l^{(t)}(\underline r, s)$, are different for different values of $l$!

If we require equation (\ref{320}) to hold for all integers $n,m$ we are implicitly assuming that all the lattice sites are occupied by an electron, so that the filling is $\nu=1$.

In \cite{mra1} we have discussed how the ONC should be rewritten in order to cover a fractional value of the filling factor. A possible way for requiring orthonormality between single electron wave functions for $\nu=\frac{1}{d}$, $d\in\N$, consists in replacing  (\ref{320}) with the following equation,
\be
\int_{\R} ds e^{isdm\frac{2\pi}{a}}\overline{h(s)}
h(s-na)=\delta_{m,0}\delta_{n,0}, \label{320bis} \en 

\noindent
for all $m,n\in \Z$,
which is obtained under the assumption that only one lattice site every $d$ is occupied. We see, therefore, that the ONC becomes very much {\em filling-dependent}, and, in a certain sense, this fact is not very surprising to us: a filling $\nu<1$ can be interpreted as if in the physical system we have less electrons than for $\nu=1$, so that these electrons are more distant from one another in order to minimize the Coulomb energy. For this reason we expect different analytic expressions for the single electron wave functions depending on the value of the filling.

We are left now with the main problem: is it possible to
produce solutions of equation (\ref{320bis})? This problem has been
partially solved in \cite{mra1} where solutions where explicitly
built up for even values of the inverse filling factor starting
from a 2-MRA and making use of the $kq$-representation. In the
next section we will show how to solve (\ref{320bis}) both for odd
and even values of $\nu^{-1}$, and this will be done without even
mentioning the $kq$-representation. However, in Section V, in
order to show the relation with our previous approach, we will
rewrite everything using the Zak transform.

\section{$d$-MRA and odd inverse filling}

In this section we  propose a method for constructing solutions of
equation (\ref{320bis}) which works for all integer values of the
inverse filling. This method produces exactly the same solutions
as in \cite{mra1} only for $\nu=\frac{1}{2}$, while produces
different wave functions for other even values  of $\nu^{-1}$.
Moreover, it will produce solutions also for those  values
of $\nu^{-1}$ which where not covered by the approach in
\cite{mra1}, that is for odd values of $\nu^{-1}$.

Let us consider a given $d$-MRA of $\Lc^2(\R)$ and its related
square-summable set of complex numbers $\{h_n\}_{n\in\Z}$
satisfying  condition (\ref{21}). Following \cite{mra1}, we use
these coefficients to define a function $T_d(\omega)$ as follows

\be   \label{41}
  T_d(s)=
  \left\{
  \begin{tabular}{ll}
  $\frac{1}{\sqrt{a}}\sum_{l\in\Z}h_le^{ils \frac{2\pi}{a}},$ &  $s\in [0,a[$ \\
  0, & otherwise.
  \end{tabular}
  \right.
\en It is clear that $T_d(s)$ is square integrable and  not
periodic. In particular, due to  condition (\ref{21}), we have
$\|T_d\|_2^2=\int_{\R}|T_d(s)|^2 ds=1$.

With this definition it is straightforward to check that \be
\int_{\R} ds e^{isdm\frac{2\pi}{a}}\overline{T_d(s)}
T_d(s-na)=\delta_{n0}\delta_{m0}. \label{42} \en Infact:

a)  the supports of $T_d(s)$ and $T_d(s-na)$ are disjoint but if
$n=0$;

b)
$$
\int_{\R} ds
e^{isdm\frac{2\pi}{a}}|T_d(s)|^2=\frac{1}{a}\sum_{l,p\in\Z}{\overline
h_l}h_p\int_0^a ds
e^{is\frac{2\pi}{a}(dm-l+p)}=\sum_{p\in\Z}h_p{\overline
h_{p+dm}}=\delta_{m0},
$$
due to condition (\ref{21}).

The physical consequences of this fact are evident: calling \be
\psi_d(x,y)=\int_{\R} K_0^{(t)}(\underline r, s)T_d(s) ds
\label{43} \en then the following ONC \be
<\psi_d,T_1^nT_2^{dm}\psi_d>=\delta_{n0}\delta_{m0}, \quad\quad
\forall n,m\in \Z, \label{44} \en holds. We must stress that our
construction does not imply that $\psi_d$ is orthogonal to
$T_1^iT_2^{j}\psi_d$ for all integer values of $i$ and $j$, but
only for those indices $(i,j)$ which belong to the set $\{(\Z,d\Z)\}$.

The set $\{T_1^{n}T_2^{dm}\psi_d(x,y): n,m\in\Z\}$ generates a triangular lattice
where not all the sites are occupied: with our choice, only a site
every $d$ along the $\vec{a_2}$ direction is non empty, while all
the sites along $\vec{a_1}$ are full. It is evident that this is
only one among many choices: another possibility is exactly the
opposite one: all the sites are occupied along $\vec{a_2}$ while
only one every $d$ is non empty along $\vec{a_1}$. This choice
corresponds to the set $\{T_1^{dn}T_2^m\psi_d(x,y): n,m\in\Z\}$.
More symmetrical choices are not difficult to imagine. Clearly,
from a mathematical point of view, all these choices are
equivalent. However, physical reasons suggest that numerical
differences could arise in the computation of the Coulomb energy,
the reason being the existence of possible different asymptotic behaviors of the
single electron wave functions for  different choices of the above
sub-lattices.   An 'a priori' decision seems very hard: the only
reasonable criteria we could use are suggested by symmetry
considerations. However we will see in Section VI that already the
choice discussed above, even if it breaks down the symmetry for
rotations of $\pi/3$ which one would expect for the triangular
Wigner crystal, produces a fairly good localized eigenfunction of
$H_0$ and, therefore, is surely worth of  a deeper analysis also
form a numerical point of view.

\vspace{3mm}

{\bf REMARKS.--} (1) It can be useful to remark that this
procedure is not unique, that is it produces as many o.n. sets in
the l-th Landau level (for any given shape of the lattice) as
many $d$-MRA we are able to produce. This is not surprising
because it reflects once more the infinite degeneracy of the
Landau levels. Of course, among all the possibilities, we are interested in finding the one which minimizes the Coulomb energy.

(2) We observe that our strategy suggests some kind of transition related to the value of $\nu$. The point is the following: in
\cite{bms} our lattice was essentially constructed with a
superposition of gaussian functions, localized around different
lattice sites. The same single-electron wave function was to be
used for any value of the filling factor, the only difference
being that the different wave functions were more or less distant
depending on $\nu$, which, as in the classical case, appears
essentially  as a  parameter fixing the distance between two
neightbouring electrons. Here the situation is more delicate.
Infact, even if the expression for $T_d$ is formally always the
same independently of the filling, see (\ref{41}), the explicit
expression changes because the coefficients $h_n$ for a, say,
2-MRA are in general different from those of a 3-MRA or, in
general, of any other $d$-MRA. We will show this in Section VI,
considering as an example the Haar 3-MRA. The consequence of this
fact is that a $d$-MRA produces an $N$-electron wave function
$\psi^{(N)}_d(\underline r_1,\underline r_2,...,\underline r_N)$
and two such functions, $\psi^{(N)}_{\frac{1}{3}}$ and
$\psi^{(N)}_{\frac{1}{5}}$ for instance, need not to have the same
analytical dependence on their variables. A natural question to be considered is the following: is there any relation between a given $\psi^{(N)}_d(\underline r_1,\underline r_2,...,\underline r_N)$ and the plateaux of the Hall resistivity corresponding to $\nu=\frac{1}{d}$? We hope to be able to consider this question in a close future.

\vspace{4mm}

As in \cite{mra1}, we show now just for completion how the
procedure can be inverted, that is how a set of coefficients
satisfying equation (\ref{21}) can be recovered by a function
satisfying the ONC (\ref{42}). The idea is quite simple: given
such a function $K_d(s)$ we simply define \be
H_n=\frac{1}{\sqrt{a}}\int_{\R} K_d(s) e^{-ins\frac{2\pi}{a}}ds,
\label{45} \en and it is now an easy computation to check that \be
\sum_{n\in\Z}H_n {\overline H_{n+dl}}=\delta_{l0}, \quad \forall
l\in \Z. \label{46} \en Indeed, using  formula
$\sum_{n\in\Z}e^{inx\frac{2\pi}{a}}=a\sum_{n\in\Z}\delta(x-na)$ we
have: \beano
&&\sum_{n\in\Z}H_n {\overline H_{n+dl}}=\frac{1}{a}\sum_{n\in\Z}\int_{\R} ds K_d(s) e^{-ins\frac{2\pi}{a}}\int_{\R} ds' {\overline K_d(s')}e^{i(n+dl)s'\frac{2\pi}{a}}=\\
&&=\sum_{n\in\Z}\int_{\R} ds' K_d(s'-na){\overline
K_d(s')}e^{idls'\frac{2\pi}{a}}. \enano Our claim now  follows
from the assumptions on $K_d$.

Finally, it is not a big surprise that whenever we take $K_d(s)$ coincident with the function
$T_d$ in (\ref{41}), then the coefficients $H_n$ in (\ref{45}) coincide with the $h_n$, $h_n=H_n$ for any $n$.

Before ending this section, it may be worth stressing that the
definitions (\ref{41}) and (\ref{45}), which are given here
without any reasonable justification, are really a consequence of
the analysis produced in \cite{mra1}, and, in particular, of the
use of the $kq$-representation, which was crucial to make evident
the relation between MRA and ONC.

\section{The role of the $kq$-representation}

In the previous section we have discussed the relation between a
$d$-MRA and an o.n. set of single electron eigestates of the free
Hamiltonian of a 2DEG in a given Landau level. Our main
requirements are that we want to generate a triangular lattice and
that we want our approach to be useful for describing those values
of the electron densities for which the plateaux are observed in
experiments. These results, which extend in a relevant way the
ones discussed in \cite{mra1}, have been obtained here without
even mentioning the $kq$-representation. Nevertheless, as we have
just remarked, this has proved to be a crucial guideline in
\cite{mra1}.  For these reasons, and because the ONC can again be
written in a much simpler form in the variables $(k,q)$, we devote
this section to restate our results by making use of the Zak
transform. Moreover, this approach will also explain why the role
of the geometry of the lattice and the particular Landau level are
completely irrelevant as far as we are only interested to the
orthonormality between the vectors $\psi$ and $T_1^nT_2^m\psi$.

We begin with adapting the definitions sketched in Section II to
our situation, that is considering the unitary operators $T_1$ and
$T_2$ defined in (\ref{315}), $T_1=e^{iQa}$ and
$T_2=e^{iP\frac{2\pi}{a}}$. A simple extension of the original
Zak's proof produces the following result:

\vspace{2mm}

{\bf Lemma 5.1}

The following set \be \psi_{kq}(p)=\sqrt{\frac{a}{2\pi}}
\sum_{n\in\Z}e^{iqna}\delta(p-k+na), \quad\quad k\in[0,a[, \quad
q\in[0,\frac{2\pi}{a}[ \label{51} \en satisfies \be
T_1\psi_{kq}(p)=e^{iqa}\psi_{kq}(p), \quad \quad
T_2\psi_{kq}(p)=e^{ik\frac{2\pi}{a}}\psi_{kq}(p), \label{52} \en
and \be \int_0^a dk \int_0^{2\pi/a}dq
\psi_{kq}(p)\overline{\psi_{kq}(p')}=\delta(p-p'). \label{53} \en
\label{lemma51}

\vspace{3mm}

This means that any $\psi_{kq}(p)$ is a common eigenstate of $T_1$
and $T_2$  for all $k$ and $q$ and, moreover, that they form a
complete set in $\Lc^2(\D)$,  $\D=[0,a[\times[0,\frac{2\pi}{a}[$,
as in Section II.

The completeness of this set can be used to write $S_{n,m}$ in a
different way. If we call $h(k,q)$ the Zaq transform of $h(s)$,
\be h(k,q)=(Zh)(k,q):=\sqrt{\frac{2\pi}{a}}\sum_{n\in
\Z}e^{-ikna}h(q+na), \label{54} \en we can find that \be
S_{n,m}=\int_\D dk dq e^{inaq+ikm\frac{2\pi}{a}}|h(k,q)|^2,
\label{55} \en which coincides essentially with the one in
\cite{mra1}. Incidentally, once more we have an evidence of the
canonicity of maps in the following sense:
$\|\psi(x,y)\|_{\Lc^2(\R^2)}=\|h(s)\|_{\Lc^2(\R)}=\|h(k,q)\|_{\Lc^2(\D)}$.
Requiring orthogonality only among $\psi_{00}$ and $\psi_{n,dm}$,
$n,m\in\Z$, we can repeat essentially the same steps as in
\cite{mra1} and we conclude that $S_{n,dm}=\delta_{n0}\delta_{m0}$
if and only if \be
J_d(k,q):=\sum_{l=0}^{d-1}|h(\frac{k+la}{d},q)|^2=\frac{d}{2\pi},
\label{56} \en which looks more friendly than 
(\ref{320bis}) because no integration appear!

The solution of this equation can be obtained starting from a set
of coefficients $\{h_n\}$ arising from a $d$-MRA, and therefore
satisfying (\ref{21}). It is enough to define the function \be
t_d(k,q)=\frac{1}{\sqrt{2\pi}}\sum_{n\in \Z}h_n
e^{ikn\frac{2\pi}{a}}, \quad (k,q)\in \D. \label{57} \en It is not
hard to check now that (1) $t_d(k,q)$ satisfies condition
(\ref{56}), and (2) the inverse Zak transform of $t_d(k,q)$
returns $T_d$ as in (\ref{41}). Indeed we have:

\beano
&&J_d(k,q):=\frac{1}{2\pi}\sum_{l,s\in\Z}h_s\overline{h_l}e^{i(s-l)\frac{2\pi k}{da}}(1+e^{i(s-l)\frac{2\pi}{d}}+e^{i(s-l)\frac{4\pi}{d}}+.......+e^{i(s-l)\frac{2(d-1)\pi}{d}})=\nonumber\\
&&=\frac{1}{2\pi}\sum_{s,p\in\Z}h_s\overline{h_{s+p}}e^{-ip\frac{2\pi
k}{da}}(\sum_{j=0}^{d-1}(e^{-ip\frac{2\pi }{d}})^j). \enano Since
$\sum_{j=0}^{d-1}(e^{ip\frac{2\pi }{d}})^j$ is equal to $d$
whenever $p=0,\pm d,\pm 2d, \pm 3d,...$ and to $0$ for all the
other values of $p$, we conclude that, with a change of variable
$m=dp$,
$$
J_d(k,q):=\frac{d}{2\pi}\sum_{s,m\in\Z}h_s\overline{h_{s+md}}e^{-im\frac{2\pi
k}{ad}},
$$
which is finally equal to $\frac{d}{2\pi}$ as a simple consequence
of equation (\ref{21}).

As for our second claim, we have \beano
&&(Z^{-1}t_d)(s)=\int_{\D}\psi_{kq}(p)t_d(k,q) dk\,dq=\\
&&=\frac{\sqrt{a}}{2\pi}\sum_{n,m\in\Z}h_n\int_0^a
dk\int_0^{2\pi/a}dq e^{iqma}
\delta(p-k+ma)e^{ink\frac{2\pi}{a}}=\\
&&=\frac{1}{\sqrt{a}}\sum_{n\in\Z}h_n\int_0^a
dk\delta(p-k)e^{ink\frac{2\pi}{a}}, \enano which is equal to zero
whenever $p\notin[0,a[$ while gives a non trivial contribution
otherwise. Computing the integral  for $p\in[0,a[$,  our assertion
easily  follows.

In analogy with our results in the previous section and with those
in \cite{mra1}, it is not hard to imagine how to use a function
$t_d(k,q)$ satisfying the ONC in its form (\ref{55}) to generate a
set of coefficients $H_n$ satisfying equation (\ref{21}),
reversing in this way the procedure and obtaining a $d$-MRA from
an o.n. set in the LLL.

\vspace{4mm}

We want to conclude this section with a remark concerning the fact
that the ONC looks identical for the square and for the triangular
lattice and also for all the  Landau levels different from the
lowest one. These facts can be nicely understood in terms of the
$kq$-representation. The idea is the following:

let $\psi$ be an eigenstate of the self-adjoint operator $H_0$
belonging to a certain eigenvalue $\epsilon$, and let us suppose
that $H_0$ commutes with the two unitary operators $T_1, T_2$
defined as in (\ref{315}). It is clear that, putting
$\psi_{nm}=T_1^nT_2^m\psi$, then $H_0 \Psi_{nm}=\epsilon
\Psi_{nm}$, for all integers $n$ and $m$. We want to answer now to
the following question:

is it possible to fix $\psi$ in such a way that \be
S_{m,n}:=<\psi,\psi_{nm}>=\delta_{n0}\delta_{m0} \label{58} \en

holds?

The answer follows quite easily now from Lemma 5.1. In fact, using the
completeness of the set $\psi_{kq}$ associated to the magnetic
translations, we can write
$$
S_{m,n}=\int_{\D}dk\,dq<\psi,\psi_{kq}><\psi_{kq},T_1^nT_2^m\psi>,
$$
and we go back to (\ref{55}), putting $h(k,q)=<\psi_{kq},\psi>$,
simply because each $\psi_{kq}$ is an eigenstate of both $T_1$ and
$T_2$. As we see, the role of the shape of the lattice, as well as
the particular Landau level we want to consider, are completely
unrelevant for this result as we have explicitly seen in Section
III.

\section{An Example: Haar for $\nu=\frac{1}{3}$}

In this section we will discuss the easiest non trivial example of
a 'trial' ground state for the 2DEG arising from a 3-MRA with an
Haar-like behavior. Let us first introduce the 3-MRA. This can be
done as follows: the characteristic function of the unit interval
$[0,1[$, $ H(x)$, is known to be a scaling function for a 2-MRA
with coefficients $h_0=h_1=\frac{1}{\sqrt{2}}$, $h_i=0$ for all
the other values of $i$. The same function can be easily shown to
be a scaling function also of a 3-MRA with coefficients
$h_0=h_1=h_2=\frac{1}{\sqrt{3}}$, $h_i=0$ for all the other values
of $i$. It is also obvious that equation (\ref{21}) is verified.
Therefore, we use this set of coefficients to construct our single
electron wave function as shown before: the function $T_3(s)$ must
be defined as
$$
T_3(s)=
  \left\{
  \begin{tabular}{ll}
  $\frac{1}{\sqrt{3a}}(1+e^{is \frac{2\pi}{a}}+e^{is \frac{4\pi}{a}}),$ &  $s\in [0,a[$ \\
  0, & otherwise.
  \end{tabular}
  \right.
$$
Using $T_3$ into (\ref{311}), together with the usual definition
of the error function, \cite{grad}, we obtain \be
\psi_3(\underline
r)=\frac{e^{\frac{i}{2}(xy-y^2/\sqrt{3})}}{\sqrt{6a\beta(1+\frac{i}{\sqrt{3}})}2\pi^{1/4}}(G(x,y,y)+G(x,y,y+\frac{2\pi}{a})+G(x,y,y+\frac{4\pi}{a})),
\label{61} \en where, as before, we have defined
$\beta=\frac{3}{8}(1-\frac{i}{\sqrt{3}})$ and \be
G(x,y,\alpha)=e^{-\frac{\alpha^2}{4\beta}-i\alpha(x-y/\sqrt{3})}\{\Phi(\frac{2\beta(x-y/\sqrt{3})-i\alpha
+2a\beta}{2\sqrt{\beta}})-
\Phi(\frac{2\beta(x-y/\sqrt{3})-i\alpha}{2\sqrt{\beta}})\}.
\label{62} \en
 $\psi_3(\underline r)$ is now used to define $(\psi_3)_{nm}(\underline r)$ as in (\ref{316}), and one can explicitly check that $$\int_{\R^2}\overline{\psi_3(\underline r)}(\psi_3)_{n3m}(\underline r)\,d^2\underline r=\delta_{n0}\delta_{m0}.$$
An interesting feature of these functions is their asymptotic
behavior which can be found considering the asymptotic behavior of
the error function, \cite{grad}. After some algebra we obtain:
\bea
&&\psi_3(\underline r)\simeq\frac{\sqrt{\beta}e^{\frac{iy}{2}(x-y/\sqrt{3})}}{\sqrt{3a}\pi^{3/4}}e^{-\beta(x-y/\sqrt{3})^2}\,\{\frac{1}{2\beta(x-y/\sqrt{3})-iy}+\nonumber\\
&&+\frac{1}{2\beta(x-y/\sqrt{3})-i(y+2\pi/a)}+\frac{1}{2\beta(x-y/\sqrt{3})-i(y+4\pi/a)}-
e^{-\beta a^2-a(2\beta(x-y/\sqrt{3})-iy)}\cdot\nonumber\\
&&\cdot[\frac{1}{2\beta(x-y/\sqrt{3})-iy+2a\beta}+\frac{1}{2\beta(x-y/\sqrt{3})-i(y+2\pi/a)+2a\beta}+\nonumber\\
&&+\frac{1}{2\beta(x-y/\sqrt{3})-i(y+4\pi/a)+2a\beta}]\}.
\label{63} \ena

What is interesting about this formula is that it shows that
whenever $x-y/\sqrt{3}\neq 0$ then $\psi_3(\underline r)$
decreases very fast while, if $x-y/\sqrt{3}= 0$, then the decay is
rather slow (like $1/|y|$ and independent of $x$). This is rather
appealing because suggests that, even if along a direction the
decay is the one expected by the Balian Low theorem,
\cite{antbag2}, outside this set of zero measure the wave function
decreases very fast {\bf in both variables}.

\vspace{3mm}

{\bf Remark.--} In \cite{mra1} we have discussed an example for
$\nu=\frac{1}{2}$, also related to the Haar basis. We give here
the result of that analysis in order to compare old and new
results and, at the same time, to correct two typos. The wave
function generating the o.n. set is \beano
T_2(x,y)=&&\!\!\!\frac{\sqrt{a}e^{-ixy/2-y^2/2}}{4\pi^{3/4}}\,(\phi(\frac{x+a-iy}{\sqrt{2}})
  -\nonumber\\
&&-\phi(\frac{x-iy)}{\sqrt{2}})+
e^{-a^2/2+ay+ixa}(\phi(\frac{x+a-i(y-a)}{\sqrt{2}})-\phi(\frac{x-i(y-a)}{\sqrt{2}})),
\enano whose asymptotic behavior can be found with the help of
\cite{grad}: \beano
T_2(x,y)\simeq&&\!\!\!\frac{\sqrt{a}e^{+ixy/2-x^2/2}}{4\pi^{5/4}}\,(\frac{1}{x-iy}+\nonumber\\
&&+\frac{1}{x-i(y-a)}
  -e^{-\pi-a(x-iy)}(\frac{1}{x+a-iy)}+\frac{1}{x+a-i(y-a)})).
\enano In particular we see that the decay of this function is
rather slow, but for a set of zero measure, when compared with
that of $\psi_3$.

\vspace{3mm}

Let us now use this wavefunction in the computation of the Coulomb
energy. We repeat the same steps as in \cite{bms}. We want to
compute the limit $\lim_{N\rightarrow
\infty}\frac{<H^{(N)}>_{\psi^{(N)}}}{N}$, where $\psi^{(N)}$ is
constructed as in (\ref{317}). We write this mean value as a sum
of two terms, a {\em kinetic} contribution coming from $H_0$,
which is very well known since all the single-electron wave
functions are eigenstates of $H_0$, and a Coulomb correction $E_c$
which is the one we really need to compute. Therefore \be
\lim_{N\rightarrow
\infty}\frac{<H^{(N)}>_{\psi^{(N)}}}{N}=\frac{\hbar
\omega}{2}+E_c, \hspace{1cm} E_c=E_W+\delta E, \label{64} \en
where $E_W$ is the classical energy per electron of the Wigner
crystal,  $E_W=-0.7821 \sqrt{\nu}$, \cite{bms}, while $\delta E$
is the quantum correction, \be \delta E=\lim_{N\rightarrow
\infty}\frac{1}{2}\sum_{(n,m)\in\C}\{[E_d((n,m),0)-E_{ex}((n,m),0)]-\frac{1}{|\underline{R}_{n,m}|}\}.
\label{65} \en Here $\C$ is the subset of $\Z\times\Z$
corresponding to the sublattice we are considering. In this case,
for instance, $\C=(\Z,3\Z)\backslash(0,0)$ to avoid self-energy
divergences. Moreover, as in Section III, we have
$\underline{R}_{n,m}=(X_{n,m},Y_{n,m})$, while the direct and the
exchange contributions are: \be
E_d((n,m),0):=\int\,d^2\underline{r}_1\,d^2\underline{r}_2
\frac{|\psi_{n,m}(\underline{r}_1)|^2|\psi(\underline{r}_2)|^2}{|\underline{r}_1-\underline{r}_2|},
\label{66} \en and \be
E_{ex}((n,m),0):=\int\,d^2\underline{r}_1\,d^2\underline{r}_2
\frac{\overline{\psi_{n,m}(\underline{r}_1)}\overline{\psi(\underline{r}_2)}\psi(\underline{r}_1)\psi_{n,m}(\underline{r}_2)}{|\underline{r}_1-\underline{r}_2|}.
\label{67} \en In this paper we are not interested in giving a
detailed result for $\delta E$,  since the CPU time required to
compute each integral is very high (more than 1 day for each
integration with a reasonable approximation). For this reason, we
only compute here the integrals using a Montecarlo method with
250000  points and restricting the integration to a large but
compact region (taking advantage from the decay behavior of
$\psi_{n,m}$ and checking first the normalization of each
wavefunction). With this very rough approximation the CPU time for each
integration is about four hours. Moreover, we  disregard  the
contributions coming from the exchange integrals, which are
usually some orders of magnitude smaller than the direct one,
\cite{bms}. Finally, due to the fact that the direct energies
$E_d((n,m),0)$ go to zero when $(n,m)$ increase, we also work with
a finite (but big) lattice.

For the above reasons we do not expect that the present result
could have any real meaning, but nevertheless it is useful because
gives a first idea of the numbers appearing in the game. The
result we have obtained is $\delta E=0.3184$ which must be
compared with the much lower result in \cite{bms}, 0.0657. It is
useful to stress that this does not imply at all that the Haar
wave-function is physically useless, but only that a different
numerical integration technique has to be used. A more detailed
numerical analysis of the procedure is now being performed. This
will include the computation of the Coulomb energy for different
values of the filling factor as well as the use of different
$d$-MRA. We hope to be able to produce physically relevant results
in a close future.

\section{Outcome}

In this paper we have carried on the analysis of the  connection
between a MRA of $\Lc^2(\R)$ and the FQHE, first proposed in
\cite{mra1}. In particular we have shown how a single electron
wave function which, together with its magnetic translates,
produces an o.n. set in the LLL can be constructed starting from a
$d$-MRA. This procedure works for $\nu=\frac{1}{d}$, $d\in\N$ and
$d\geq 2$. We have also shown that this procedure can be
essentially inverted since to any o.n. basis of translated
functions of the LLL (corresponding to $\nu=\frac{1}{d}$)
corresponds a set of coefficients satisfying the main condition of
a $d$-MRA of $\Lc^2(\R)$.

All of these results have been obtained working with a triangular
lattice, which is the one suggested by classical and quantum
energetic considerations. The extension to higher Landau levels has
also been briefly outlined.

We have also given an example of our construction for
$\nu=\frac{1}{3}$, using the Haar 3-MRA, and for the related o.n.
basis in the configuration space we have also obtained an
approximated value of the Coulomb energy.

It is clear that our method produces a wide set of possible o.n.
functions in the LLL, and our future interest consists mainly in going
beyond the simple Haar choice trying to minimize the Coulomb
energy so to recover completely the experimental data.

As already mentioned, it will be very interesting also to analyze the (possible) connection between the different wave functions $\psi_d^{(N)}$ produced by our method and the plateux corresponding to $\nu=\frac{1}{d}$. Of course, this mechanism is completely {\em non-standard}, so that an analysis of the relations between our approach and the one originated by Laughlin should also be worked out.

\section*{Acknowledgements}
I would like to thank Prof. M. Zarcone, Prof. F. Oliveri, Prof. G.
Russo and Dr. F. Agostaro for their suggestions concerning the
numerical computation in Section VII, and Prof. M. Sammartino for
giving me most of his CPU time (for free!!!).

I also would like to acknowledge financial support by the Murst,
within the  project {\em Problemi Matematici Non Lineari di
Propagazione e Stabilit\`a nei Modelli del Continuo}, coordinated
by Prof. T. Ruggeri.

\end{document}